\documentclass[11pt,a4paper]{article}
\pdfoutput=1

\usepackage{jheppub}
\usepackage{latexsym}
\usepackage{revsymb}
\usepackage{multirow}
\usepackage{color}
\usepackage[usenames,dvipsnames,svgnames,table]{xcolor}

\usepackage{graphicx}% Include figure files
\usepackage{epsfig}  % Include figure files
\usepackage{epsf}    % Include figure files
\usepackage{dcolumn}% Align table columns on decimal point
\usepackage{bm}% bold math
\usepackage{dcolumn}% Align table columns on decimal point
\usepackage{textcomp}% Align table columns on decimal point
\usepackage{float}
\usepackage{subfig}
\usepackage{hypcap}
\usepackage[]{hyperref}

\hypersetup{
  bookmarks=true,         % show bookmarks bar?
  unicode=false,          % non-Latin characters in Acrobat?s bookmarks
  pdftoolbar=true,        % show Acrobat?s toolbar?
  pdfmenubar=true,        % show Acrobat?s menu?
  pdffitwindow=true,     % window fit to page when opened
  pdfstartview={FitH},    % fits the width of the page to the window
  pdfsubject={Neutrino Oscillations Phenomenology},   % subject of the document
  pdfnewwindow=true,      % links in new window
  pdfcreator={RevTeX},
  colorlinks=true,       % false: boxed links; true: colored links
  linkcolor=red,          % color of internal links
  citecolor=blue,        % color of links to bibliography
  filecolor=black,      % color of file links
  urlcolor=blue,           % color of external links
}

\def\anu{{\bar\nu}}

\newcommand{\bma}{\boldmath}
\newcommand{\ubma}{\unboldmath}
\newcommand{\beq}{\begin{equation}}
\newcommand{\eeq}{\end{equation}}
\newcommand{\beqa}{\begin{eqnarray}}
\newcommand{\eeqa}{\end{eqnarray}}

\newcommand{\ty}{{\theta_{13}}}
\newcommand{\tz}{{\theta_{23}}}

\newcommand{\dcp}{\delta_{\mathrm{CP}}}

\newcommand{\pme}{P_{\mu e}}

\newcommand{\dxx}{\Delta\chi^2}

\newcommand{\sig}{$\sigma~$}
\newcommand{\dmm}{\Delta m^2_{\mu\mu}}
\newcommand{\dee}{\Delta m^2_{ee}}
\newcommand{\sdmm}{\sigma(|\Delta m^2_{\mu\mu}|)}

\begin{document}

\DeclareGraphicsExtensions{.eps,.ps}

\title{High-precision measurement of 
atmospheric mass-squared splitting with T2K and NOvA}

\author[a]{Sanjib Kumar Agarwalla,}
\author[b]{Suprabh Prakash,} 
{\author[c]{Wei Wang$\,$}

\affiliation[a]{Institute of Physics, Sachivalaya Marg, Sainik School Post, Bhubaneswar 751005, India}
\affiliation[b]{Harish-Chandra Research Institute, Chhatnag Road, Jhunsi, Allahabad 211019, India}
\affiliation[c]{Physics Department, College of William and Mary, Williamsburg, VA}

\emailAdd{sanjib@iopb.res.in}
\emailAdd{suprabhprakash@hri.res.in}
\emailAdd{wswang@wm.edu}

\begin{abstract}
{A precise measurement of the atmospheric mass-squared splitting
$|\dmm|$ is crucial to establish the three-flavor paradigm and to
constrain the neutrino mass models. In addition, a precise value of
$|\dmm|$ will significantly enhance the hierarchy reach of future
medium-baseline 
reactor experiments like JUNO and RENO-50. In this work, we explore the
precision in $|\dmm|$ that will be available after the full runs of T2K and 
NO$\nu$A. We find that the combined data will be able to improve the
precision in $|\dmm|$ to sub-percent level for maximal 2-3 mixing. Depending
on the true value of $\sin^2\tz$ in the currently-allowed
3\sig range, the precision in $|\dmm|$ will vary from 0.87\% to 1.24\%. 
We further demonstrate that this is a robust measurement as it remains almost
unaffected by the present uncertainties in $\ty$, $\dcp$, the choice of mass
hierarchy, and the systematic errors.
}
\end{abstract}


%\pacs{14.60.Pq,14.60.Lm,13.15.+g}
\preprint{IP/BBSR/2013-18}

\maketitle

%%%%%%%%%%%%%%%%%%%%%%%%%%%%%%%%%%%%%%%%%%%%%%%%%%%%%%%%%%%%%%%%%%%%%%%%%%%%%%%%%%%%%%%%%%%
%%%%%%%%%%%%%%%%%%%%%%%%%%%%%%%%%%%%%%%%%%%%%%%%%%%%%%%%%%%%%%%%%%%%%%%%%%%%%%%%%%%%%%%%%%%
%%%%%%%%%%%%%%%%%%%%%%%%%%%%%%%%%%%%%%%%%%%%%%%%%%%%%%%%%%%%%%%%%%%%%%%%%%%%%%%%%%%%%%%%%%%

\section{Introduction}
Recent discovery of a moderately large value of
$\ty$~\cite{An:2012eh,An:2012bu,An:2013zwz,Ahn:2012nd}
%, quite close to its
%previously-suggested 3\sig upper bound~\cite{}, has brought 
has provided an edge for
the present generation long-baseline superbeam experiments 
to explore the remaining fundamental unknowns
like neutrino mass hierarchy (MH), octant of
$\tz$ and the leptonic CP-violation.
%Neutrino oscillation experiments are entering into
%the regime of measuring the era
%Dirac $CP$ phase, determining the $\theta_{23}$ octant and resolving mass
%hierarchy~(MH) MH, theta_23 octant, and CP
%after the last mixing angle $\theta_{13}$ has been measured
%by the current generation short-baseline reactor neutrino experiments
%represented by Daya Bay~\cite{An:2012eh,An:2012bu,An:2013zwz,Ahn:2012nd}. 
T2K~\cite{Itow:2001ee,Abe:2011ks} and
NO$\nu$A~\cite{Ayres:2002ws,Ayres:2004js,Ayres:2007tu,Patterson:2012zs}
are the two current generation experiments that have potential to shed light
on these remaining unknowns using the $\theta_{13}$ driven $\nu_\mu
/\bar{\nu}_{\mu} \rightarrow \nu_e/\bar{\nu}_e$ appearance
channel~\cite{Hagiwara:2006nn,Huber:2009cw, Agarwalla:2012bv,
  Prakash:2012az, Chatterjee:2013qus, Blennow:2013swa, Agarwalla:2013ju,
  Choubey:2013xqa}. 
Another important consequence of 
the large value of $\theta_{13}$ is that it has enabled 
the medium-baseline reactor oscillation~(MBRO) experiments 
like JUNO \cite{Yifang:2012} and
RENO-50 \cite{RENO-50} to resolve MH
~\cite{Petcov:2001sy,Learned:2006wy,Zhan:2008id,Zhan:2009rs,Qian:2012xh,
Ge:2012wj,Li:2013zyd,Kettell:2013eos,Capozzi:2013psa}. 
%A clear
%determination of MH will shed light to neutrino mass
%models and provide valuable information to future neutrinoless double beta
%decay experiments. 
While it is important for T2K and NO$\nu$A to address these pressing issues,
it has been pointed out in 
\cite{deGouvea:2005hk, Nunokawa:2005nx, Qian:2012xh, Li:2013zyd,
  Kettell:2013eos} that the sensitivity of MBRO experiments to MH can be
significantly improved by a high-precision measurement of $|\dmm|$.
T2K and NO$\nu$A can do this measurement via the  $\nu_\mu
/\bar{\nu}_{\mu} \rightarrow \nu_\mu/\bar{\nu}_{\mu}$ disappearance channel,
\beq
\label{eq:survival}
P(\nu_\mu/\bar{\nu}_{\mu} \rightarrow \nu_\mu/\bar{\nu}_{\mu}) = 1 -
\sin^22\theta_{\mu\mu}\sin^2\left(\frac{\dmm L}{4E}\right).
\eeq
Here $|\dmm|$ and $\theta_{\mu\mu}$ are the effective two-flavor atmospheric
mass-squared splitting and mixing angle, measured in muon neutrino disappearance
oscillation experiments~\cite{Nunokawa:2005nx,deGouvea:2005hk}\footnote{For the 
experiments under consideration, $\Delta m^2_{21}L/E \ll 1$
(where $\Delta m^2_{ij}=m^2_i-m^2_j$)
and can be treated as a small perturbation in obtaining Eq. 
\ref{effective-atmospheric-splitting} and \ref{effective-mixing}.},
\beqa
\label{effective-atmospheric-splitting}
\Delta m^2_{\mu\mu} = \Delta m^2_{31} &-& \Delta m^2_{21} (\cos^2 \theta_{12}
- \cos \dcp \sin \theta_{13} \sin 2 \theta_{12} \tan\theta_{23}), \\
\label{effective-mixing}
\sin^2 2\theta_{\mu\mu} &=& 4 \cos^2\theta_{13}\sin^2
\theta_{23}(1-\cos^2\theta_{13}\sin^2\theta_{23}).
\eeqa
%The value of $\Delta m^2_{31}$ is calculated separately for NH and for IH
%using this equation where $\Delta m^2_{\mu\mu}$ is taken to be positive for
%NH and negative for IH.
On the one hand, precision in $|\dmm|$ can mitigate the challenge in the
absolute energy scale uncertainty in MBRO experiments, thus enhancing their
sensitivity to MH. On the other hand, comparison of the effective $|\dmm|$
from muon-flavor oscillation experiments and the corresponding effective
$|\dee|$ from electron-flavor oscillation experiments can provide additional
MH information
~\cite{Nunokawa:2005nx,deGouvea:2005hk,Minakata:2007tn,Zhang:2013rta}.
Recently, it has been demonstrated that a precision of 1\% on $|\dmm|$ can
improve JUNO's sensitivity to MH from $\Delta \chi^2 = 10$ to $\Delta \chi^2
= 18$ in a six-years run~\cite{Li:2013zyd}.
%Ref.~\cite{Zhang:2013rta} points out the current global data
%have a hint of normal hierarchy~(NH) at 1-$\sigma$ confidence
%level~(C.L.). 
Besides addressing the need of MBRO experiments, a precise $|\dmm|$
measurement, along with a precision measurement of $|\dee|$,
is a crucial step towards validating the 3-flavor oscillation
model~\cite{Nunokawa:2005nx,Zhang:2013rta}. An accurate $|\dmm|$ measurement
will also severely constrain the neutrino mass models~\cite{Albright:2006cw}
and itself a key input for neutrinoless double beta decay
searches~\cite{Pascoli:2005zb}.

Currently the most precise information on $|\dmm|$ comes 
from the MINOS experiment. A two-flavor analysis
based on its complete run gives
$|\dmm|=2.41^{+0.09}_{-0.10}\times10^{-3}$ eV$^2$~\cite{Adamson:2013whj},
which corresponds to a relative 1$\sigma$ precision of $\sigma(\Delta
m^2_{\mu\mu})=3.94\%$\footnote{We define the relative 1$\sigma$ error as
1/6th of the $\pm 3\sigma$ variations around the best-fit.}.
The latest disappearance analysis from T2K experiment based on its
3.86\% of the total exposure, {\it i.e.} $3.01\times10^{20}$ protons on
target~(p.o.t), gives $|\Delta
m^2_{32}|=2.44^{+0.17}_{-0.15} \times
10^{-3}$~\text{eV}$^2$~\cite{Abe:2013fuq}\footnote{The T2K result adapts a
  three-flavor analysis and the quoted number assumes normal MH.}.
The current T2K precision is only $\sigma(\Delta m^2_{32})=6.56\%$. 
In this paper, we explore whether it is plausible to reach the 1\% precision 
with the combined data from T2K and NO$\nu$A. 
These two experiments will gather 
copious statistics from the muon disappearance channel,
enabling a high-precision measurement of $\dmm$.
In Sec.~\ref{exp_spec}, we
briefly mention the key experimental features of T2K and NO$\nu$A
and provide the simulation details adapted in this work. In Sec.~\ref{results}, we
discuss the precision in $|\dmm|$ achievable by these two experiments and
its dependence on various factors. Finally, we give our concluding remarks
in Sec.~\ref{sec:summary}.
%Our definition of $\dmm$ is in App.~\ref{app:dmm}. 

%On the other hand, long-baseline neutrino oscillation
%experiments like NO$\nu$A and T2K are directly measuring the effective
%mass-squared splitting $\Delta m^2_{\mu\mu}$, which contains the MH
%information. 

%%%%%%%%%%%%%%%%%%%%%%%%%%%%%%%%%%%%%%%%%%%%%%%%%%%%%%%%%%%%%%%%%%%%%%%%%%%%%%%%%%%%%%%%%%%
%%%%%%%%%%%%%%%%%%%%%%%%%%%%%%%%%%%%%%%%%%%%%%%%%%%%%%%%%%%%%%%%%%%%%%%%%%%%%%%%%%%%%%%%%%%
%%%%%%%%%%%%%%%%%%%%%%%%%%%%%%%%%%%%%%%%%%%%%%%%%%%%%%%%%%%%%%%%%%%%%%%%%%%%%%%%%%%%%%%%%%%

\section{Experimental specifications and simulation details}\label{exp_spec}

%In this section, we briefly describe the two 
%present-generation experiments that we have 
%considered in this work, namely, T2K and NO$\nu$A.

\subsection{The T2K experiment}

The Tokai to Kamioka (T2K) experiment is 
%presently running 
%located in Japan and is 
the first experiment to observe the three
flavor effects in neutrino oscillations and its main objective is 
to measure $\ty$ by observing
$\nu_{\mu}/\bar{\nu}_{\mu}\rightarrow\nu_{e}/\bar{\nu}_e$ oscillations. 
Neutrinos are produced in the J-PARC
accelerator facility in Tokai and are directed towards
the 22.5 kton water \v{C}erenkov Super-K detector placed 
in Kamioka, 295 km away at a $2.5^\circ$ off-axis angle \cite{Itow:2001ee}. 
%For the muon events 
For muon charged-current quasi-elastic~(CCQE) events,
the energy resolution is $\sigma_{E}(\text{GeV})=0.075\sqrt{E/\text{GeV}}+0.05$.
The $\nu_{\mu}$ beam peaks sharply at 0.6 GeV, which is
very close to the 1st oscillation maximum of $\pme$.
The flux falls off rapidly, such that, there is hardly
any at energies greater than 1 GeV. The experiment
plans to run with a proton beam power of 750 kW
with proton energy of 30 GeV for 5 years in $\nu$
mode only. This corresponds to a total exposure of
$8\times10^{21}$ protons on target~(p.o.t). The neutrino
flux is monitored by the near detectors, located 280 m away from the point
of neutrino production. The background information 
and other details are taken from references \cite{fechnerthesis,Huber:2009cw}.

\subsection{The NO{\bma$\nu$\ubma}A experiment}

The NO$\nu$A~(NuMI\footnote{Neutrinos at the Main Injector.} 
Off-axis $\nu_{e}$ Appearance) experiment
\cite{Ayres:2007tu,Patterson:2012zs,Childress:2013npa}
uses FermiLab's NuMI $\nu_{\mu}/\bar{\nu}_\mu$ beamline and is
scheduled to start taking data from late 2013.
A 14 kton Totally Active Scintillator Detector (TASD)
will be placed in Ash River, Minnesota which is
810 km away at an off-axis angle of 14 mrad ($0.8^\circ$).
This off-axis narrow-width beam peaks at 2 GeV.
A 0.3 kton near detector will be located at the FermiLab site to monitor the
un-oscillated neutrino or anti-neutrino flux.
It aims to determine the unknowns such as MH,
leptonic CP-violation, $\ty$ and the octant of $\tz$ by
the measurement of $\nu_{\mu}/\bar{\nu}_{\mu}\rightarrow\nu_{e}/\bar{\nu}_e$
oscillations. For the CCQE muon events, the energy resolution is
%a gaussian with 
$\sigma_{E}(\text{GeV})=0.06\sqrt{E/\text{GeV}}$.
The experiment is scheduled to run for 3 years in
$\nu$ mode followed by 3 years in $\anu$ mode
with a NuMI beam power of $0.7$ MW and 120 GeV
proton energy, corresponding to $6\times 10^{20}$ p.o.t per year.

\subsection{Simulation details}
\label{simulation}

%Below, we describe the details of the simulations.
We use GLoBES~\cite{Huber:2004ka,Huber:2007ji} 
to carry out all the simulations in this work. The true values
of neutrino oscillation parameters have been taken to be:
$\Delta m^2_{21} = 7.5 \times  10^{-5}$~\text{eV}$^2$, 
$\sin^2 \theta_{12} = 0.3$~\cite{GonzalezGarcia:2012sz,NuFIT},
$|\Delta m^2_{\mu\mu}| = 2.41 \times 10^{-3}$~\text{eV}$^2$~\cite{Adamson:2013whj,Nichol:2013caa}, 
and $\sin^2 2 \theta_{13} =
0.089$~\cite{An:2012eh,Ahn:2012nd,Abe:2011fz,Abe:2012tg}.
$\Delta m^2_{31}$ is calculated based on $\dmm$ and other values using
Eq.~\ref{effective-atmospheric-splitting} assuming different true MH and
$\dcp$.
The value of $\Delta m^2_{31}$ is calculated separately 
for normal hierarchy~(NH where $m_3>m_2>m_1$) and for inverted hierarchy~(IH
where $m_2>m_1>m_3$) using this equation where $\Delta m^2_{\mu\mu}$ is
taken to be +ve for NH and -ve for IH. We have taken into account the
present 3\sig uncertainty of $\sin^2\tz$ in the range
0.36 to 0.66~\cite{GonzalezGarcia:2012sz, NuFIT} both in simulated data and in
fit. Note that, we perform a full
three-flavor analysis in obtaining the results.
{\it We find that the true value of $\dcp$ has little impact 
to the precision of $|\dmm|$. Therefore, in this work, $\dcp\text{(true)}=0$
has been assumed for all the results.} 
%We demonstrate this effect in appendix \ref{appB}.
The experimental features of T2K and re-optimized NO$\nu$A are the same as
considered in reference~\cite{Agarwalla:2012bv}.
We consider the nominal set of systematics i.e. normalization error of 2.5\% and 10\%
on signal and background respectively for both the experiments.
We also consider the tilt error\footnote{Here ``tilt'' describes a linear
distortion of the event spectrum.} on signal and backgrounds
to incorporate the energy-scale uncertainty.
In this work, we consider 0.01\% tilt error for NO$\nu$A
and 0.1\% tilt error for T2K, for both signal and backgrounds. 
The impact of different assumptions on systematics has been studied further 
in Sec.~\ref{systematics}.

Fig.~\ref{events_pmm} shows the survival event
spectra~(CCQE muon events) for T2K and NO$\nu$A for three different
choices of $|\dmm|$. These three different choices correspond to the
best-fit and the 3\sig upper and lower bounds.
%A distiction can be made on the basis of total number of events also. 
The total events corresponding to $|\dmm|=2.1\times10^{-3}$~\text{eV}$^2$,
$2.41\times10^{-3}$ eV$^2$, and $2.7\times10^{-3}$ eV$^2$
are 230, 153, and 114 respectively, with a three-years $\nu$ run in
NO$\nu$A. The corresponding numbers for T2K are 369, 300, and 318 with a
five-years $\nu$ run. The upper left (right) panel shows the event spectrum
for the experiment T2K (NO$\nu$A). The
ratio of oscillated to un-oscillated event spectrum are give in the lower
panels. Fig.~\ref{events_pmm} shows that the first oscillation minima are
clearly seen in both experiments due to their excellent energy resolution
for CCQE muon events. This enables them to perform an accurate 
measurement of $|\dmm|$.

\begin{figure}[H]
\centering
\includegraphics[width=0.9\textwidth]
{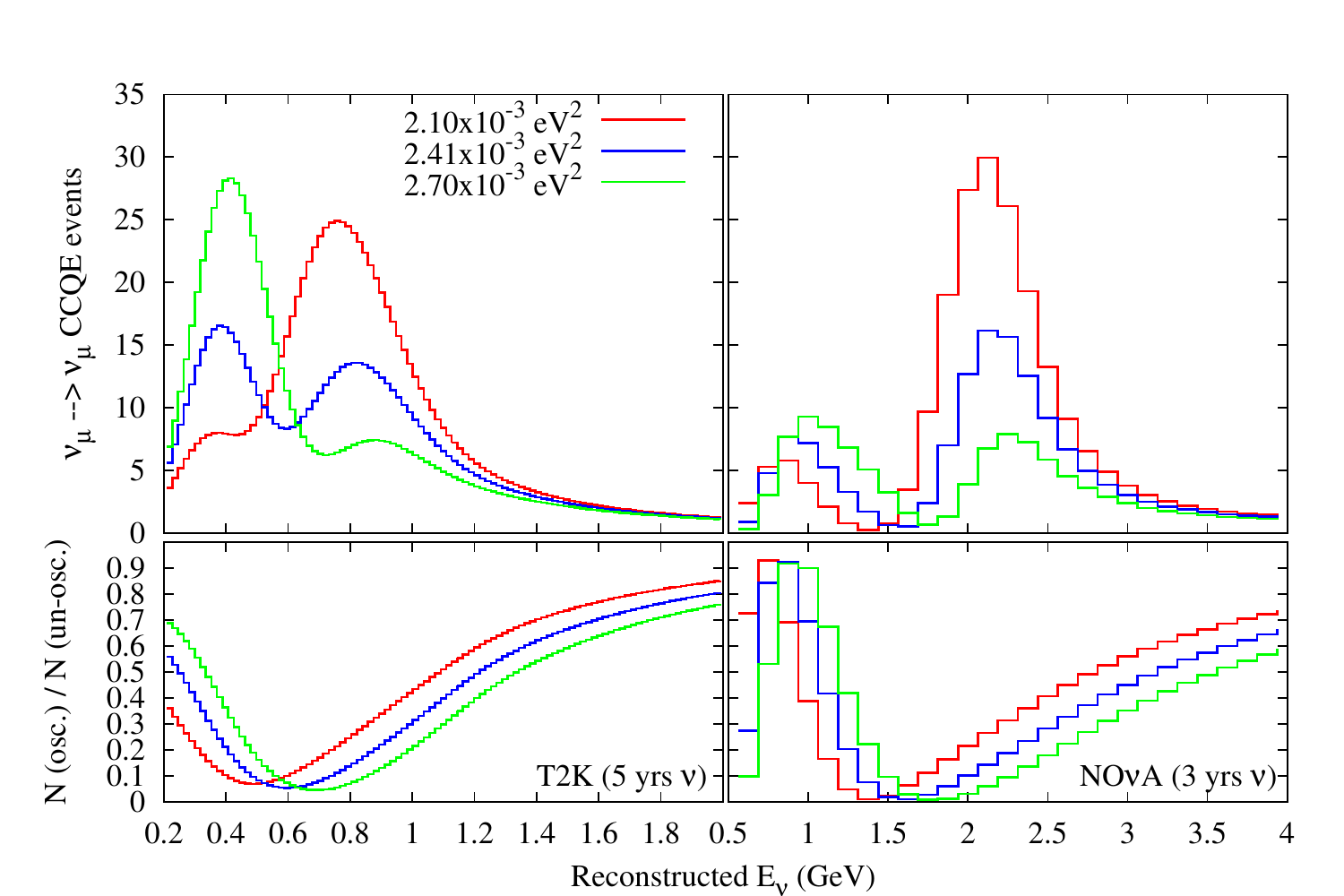}
\caption{\footnotesize{Reconstructed event spectrum for the experiments T2K (left panels)
  and NO$\nu$A (right panels) for the three different values of $|\dmm|$
  corresponding to the present best-fit and upper and lower $3\sigma$
  limits. Only CCQE $\nu_{\mu}$ survival events have been considered. The
  top panels show the event spectra while the bottom panels show the
  ratio of oscillated over un-oscillated events as a function of the
  reconstructed energy. We have assumed NH, $\sin^2\tz=0.5$,
  $\sin^22\ty=0.089$, and $\dcp=0$.}}
\label{events_pmm}
\end{figure}

The precision of $|\dmm|$ is calculated using the conventional least
chi-squared method. To
calculate the $\dxx$, the observed number of events are
simulated using a particular choice of the true parameters.
These are then contrasted with the events generated using 
another test set of oscillation parameters. This procedure
is repeated for all the test values of oscillation 
parameters in their respective allowed intervals.
We marginalize over test $\sin^22\ty$ in its
2\sig range, over test $\dcp\in[-180^\circ,180^\circ]$ and 
over test $\sin^2\tz$ in the 3\sig range. We impose a Gaussian
prior in $\sin^22\ty$ with 5\% uncertainty~\cite{TheDayaBay:2013kda}. The
solar parameters are kept fixed; and so is the Earth matter density. GLoBES
performs a binned-spectral analysis using a Poissonian definition of the 
$\dxx$. 
The relative 1$\sigma$ precision of $|\dmm|$ is defined as 
\beq 
\sdmm = \frac{(|\dmm|)^{+3\sigma}-(|\dmm|)^{-3\sigma}}{6}\times
\frac{100}{2.41\times10^{-3}~\text{eV}^2} \%,
\eeq 
where $2.41\times10^{-3}~\text{eV}^2$ is the present best-fit of $|\dmm|$. 
$(|\dmm|)^{+3\sigma}$ and $(|\dmm|)^{-3\sigma}$
are the two values of $|\dmm|$ at which $\dxx=9$; with $(|\dmm|)^{+3\sigma}$ being the
larger of the two.
%The reason we have chosen to use the 3$\sigma$ range to identify the
%1$\sigma$ precision is because $\dxx$ is generally more stable at 3$\sigma$.
%%%%%%%%%%%%%%%%%%%%%%%%%%%%%%%%%%%%%%%%%%%%%%%%%%%%%%%%%%%%%%%%%%%%%%%%%%%%%%%%%%%%%%%%%%%
%%%%%%%%%%%%%%%%%%%%%%%%%%%%%%%%%%%%%%%%%%%%%%%%%%%%%%%%%%%%%%%%%%%%%%%%%%%%%%%%%%%%%%%%%%%
%%%%%%%%%%%%%%%%%%%%%%%%%%%%%%%%%%%%%%%%%%%%%%%%%%%%%%%%%%%%%%%%%%%%%%%%%%%%%%%%%%%%%%%%%%%

\section{Study of the $|\dmm|$ precision}
\label{results}
%In this section, we discuss the results that we have obtained. 
In the following subsections, we study the effect of various important
issues like contribution from appearance channel, 
the effect of uncertainty in $\sin^2\tz$ and the 
effect of difference systematic uncertainties, on the
precision of $\sdmm$. Finally, we show how the precision of $\sdmm$ is 
going to improve with increasing statistics from these two experiments. 
%In appendix \ref{appA}, we show the effect of the uncertainty of
%$\sin^22\ty$ on $\sdmm$ based on the current experimental data. 

\subsection{Effect of appearance and disappearance data}

%In this subsection, we show the relative contribution from the
%appearance and disappearance channels to the accuracy in $\dmm$. 
Fig.~\ref{dmm_appdiapp} shows the $\dxx$ vs. test $|\dmm|$
for the NO$\nu$A experiment, assuming NH (IH) to be the true hierarchy in the left (right) panel, 
$\sin^2\tz\text{(true)}=0.5$ and
$|\dmm|\text{(true)}=2.41\times10^{-3}~\text{eV}^2$.
All test parameters have been marginalized over, except 
the solar parameters as we explained earlier. It can be seen from
Fig. \ref{dmm_appdiapp} that the precision is dominated by the
disappearance data. The combined data of disappearance and appearance channels
improves the precision by 0.04\%, compared to disappearance alone.
The contribution of appearance channel to the determination
of $|\dmm|$ is very small. For completeness, we still include
both appearance and disappearance data in this work.

\begin{figure}[H]
\centering
\includegraphics[width=0.49\textwidth]
{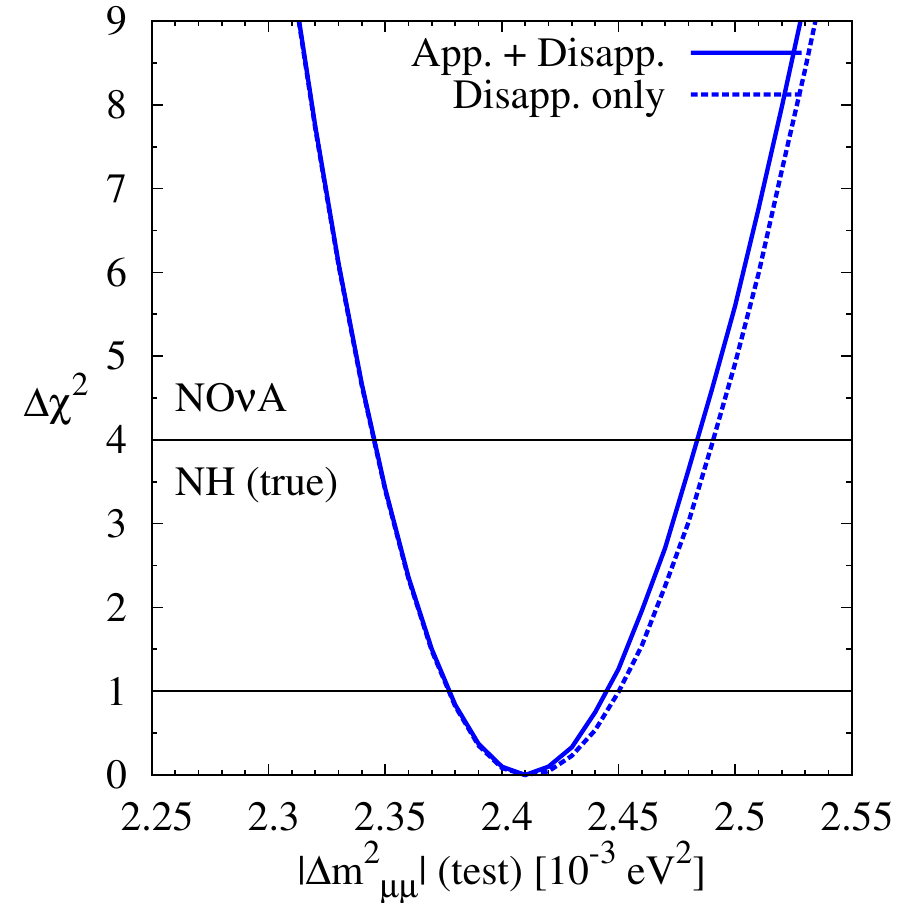}
\includegraphics[width=0.49\textwidth]
{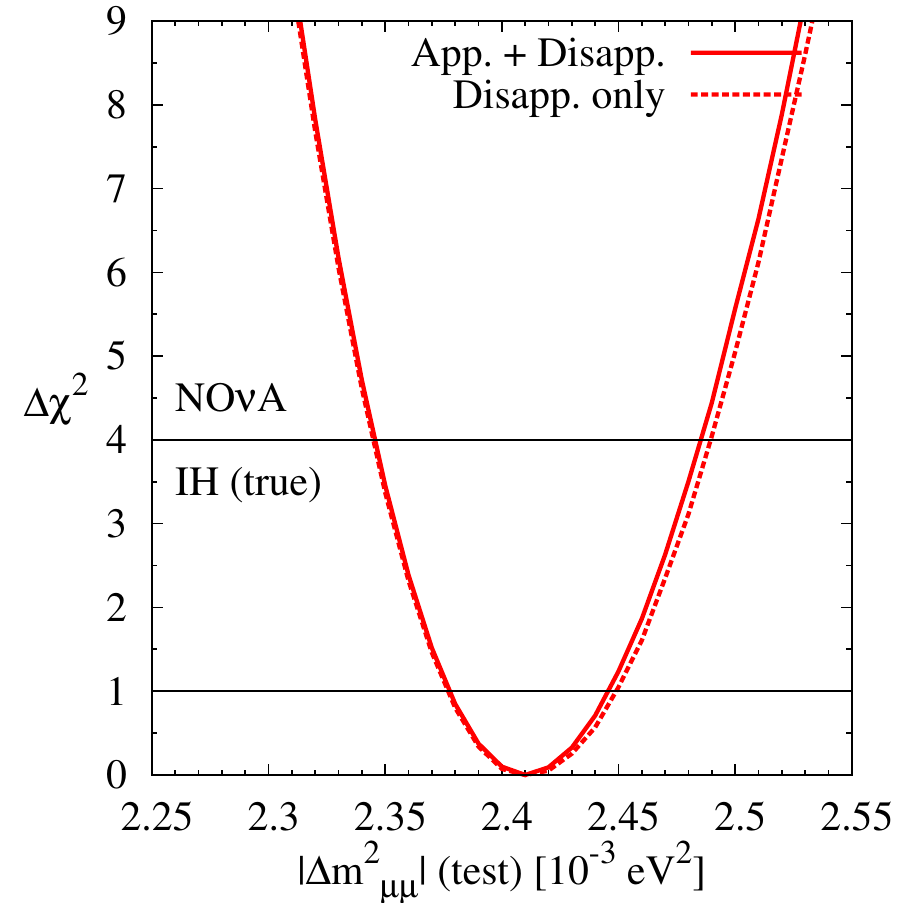}
\caption{\footnotesize{$\dxx$ vs. test $|\dmm|$ for NO$\nu$A. Left (Right) panel
  corresponds to NH (IH) being the true hierarchy. The relative contribution
  to the sensitivity from disappearance and appearance channels is
  shown. Here we take $|\dmm|=2.41\times10^{-3}$ eV$^2$, $\sin^2\tz
  \text{(true)}=0.5$, $\sin^22\ty\text{(true)}=0.089$, and
  $\dcp\text{(true)}=0$.}}
\label{dmm_appdiapp}
\end{figure}

\subsection{Precision of $|\dmm|$ with T2K and NO$\nu$A}
%$\Delta\chi^2$ vs. test $|\Delta m^2_{\mu \mu}|$ }

In Fig. \ref{dmm_allexp}, we compare the precision of the two 
experiments T2K and NO$\nu$A in measuring $|\dmm|$. The
left (right) panel corresponds to NH (IH)
being the true hierarchy. As before, we assume $\sin^2\tz\text{(true)}=0.5$. 
We find that, after full runs, NO$\nu$A will give $\sdmm=1.45\%$, while T2K
will give a more precise measurement $\sdmm=1.16\%$. The reason
is that T2K has more statistics. 

\begin{figure}[H]
\centering
\includegraphics[width=0.49\textwidth]
{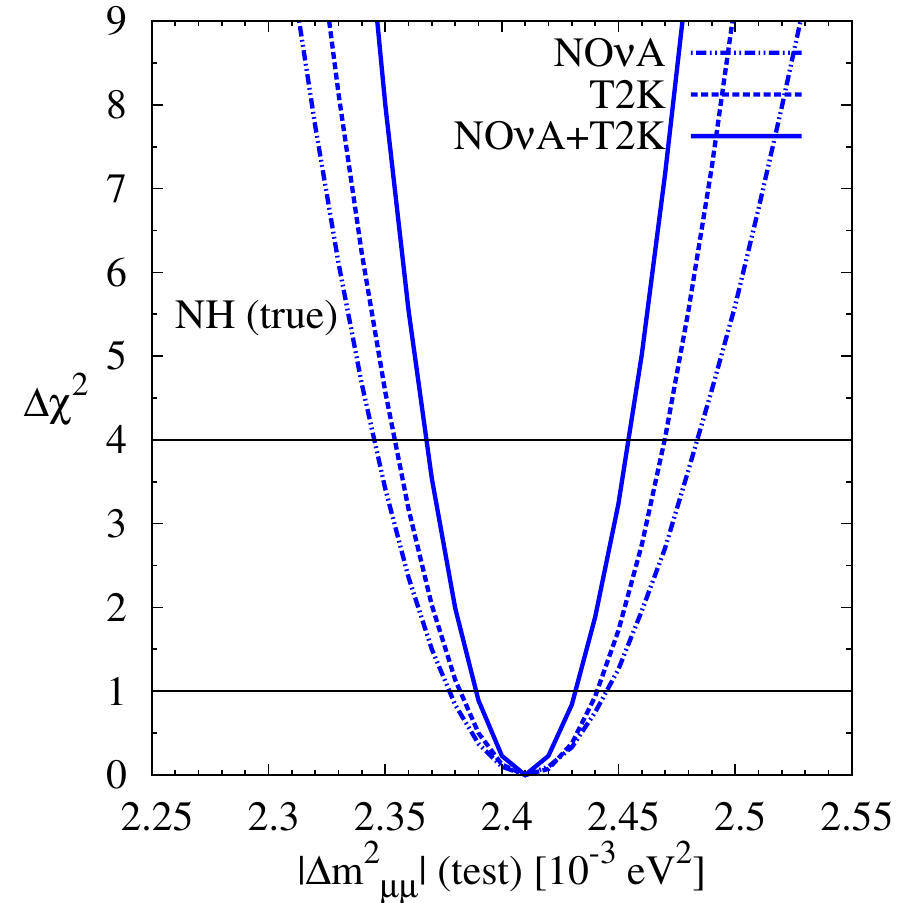}
\includegraphics[width=0.49\textwidth]
{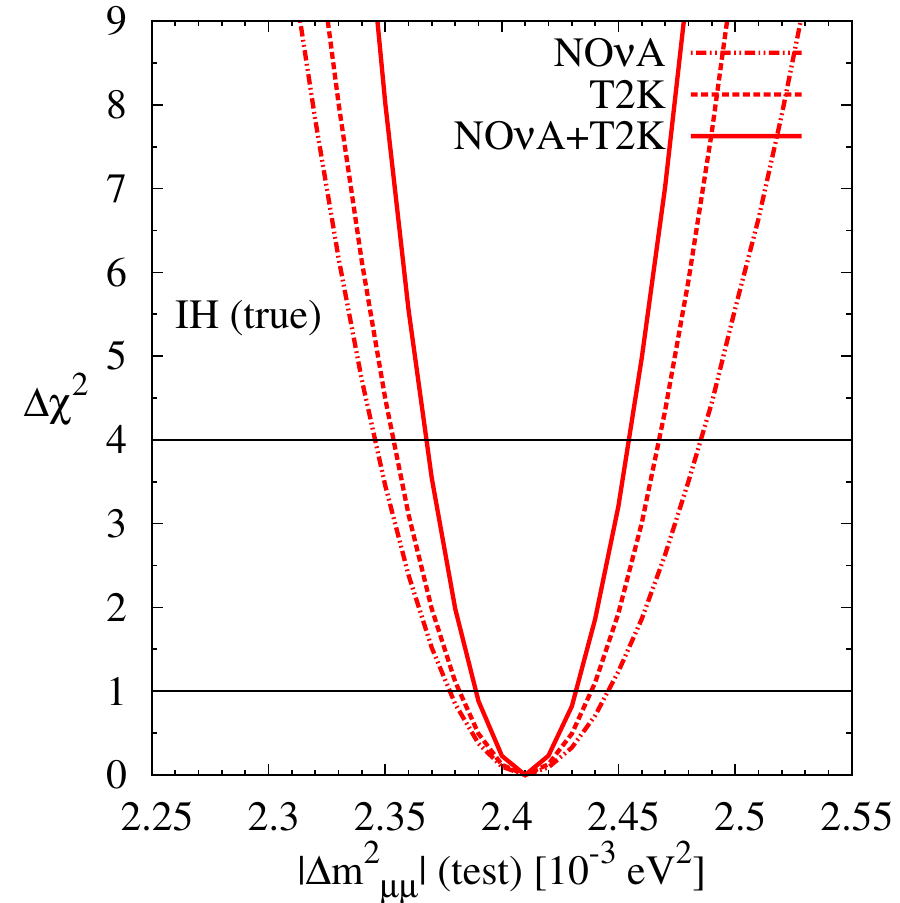}
\caption{\footnotesize{$\dxx$ vs. test $|\dmm|$ for
T2K and NO$\nu$A alone and the combined data. Left (Right) panel corresponds
to NH (IH) being the true hierarchy.
Here $|\dmm|\text{(true)}=2.41\times10^{-3}$ eV$^2$,
$\sin^2\tz\text{(true)}=0.5$, $\sin^22\ty\text{(true)}=0.089$, and
$\dcp\text{(true)}=0$.}}
\label{dmm_allexp}
\end{figure}
We next explore the potential of combined data. A precision of
$\sdmm=0.87\%$ can be obtained after the full runs of these two experiments.
{\it Thus, if the 2-3 mixing is maximal, then a less than 1\% accurate
determination of $|\dmm|$ can be achieved.}

It can also be seen from the $\dxx$ vs. test $|\dmm|$ plots that
the precision of $|\dmm|$ is essentially independent of the hierarchy. Thus
for simplicity, from here onwards, we show results only for NH assumed to be
the true hierarchy.

\subsection{Impact of 2-3 mixing angle on $|\Delta m^2_{\mu \mu}|$ precision}

Recent MINOS results hint at a non-maximal $\sin^22\tz$~\cite{Adamson:2013whj}.
Global analysis~\cite{GonzalezGarcia:2012sz,NuFIT} suggests two
degenerate values of $\tz$, one in the lower octant 
and the other in the higher octant. The leading
term in the muon disappearance probability is dependent on $\sin^2\tz$ 
as shown in Eq.~\ref{effective-mixing}. Thus, this parameter
is expected to affect the precision in the measurement of $|\dmm|$ directly.
In this section, we study the dependence of $|\dmm|$ precision on the true
value of $\sin^2\tz$. 

Fig. \ref{dmm_theta23} shows the effect of $\sin^2\tz$ on the determination
of $|\dmm|$. The left panel shows the 3\sig allowed regions in the 
$|\dmm|\text{(test)}$ - $\sin^2\tz\text{(true)}$ plane for the experiments
T2K, NO$\nu$A, and combined. The right panel depicts the corresponding relative
1$\sigma$ precision on $|\dmm|$. It can be seen that 
the best precision can be achieved for the maximal 2-3 mixing case and it
deteriorates as the mixing deviates from maximal.
%$\sdmm$
%is the least when $\sin^2\tz=0.5$ and it increases as the mixing deviates
%from maximal. 
With the combined data of T2K and NO$\nu$A, a $\sdmm=0.87\%$
is achievable for $\sin^2\tz\text{(true)}=0.5$. For the most conservative
choice of $\sin^2\tz\text{(true)}=0.36~\textrm{or}~0.66$ at the 3$\sigma$
allowed limits, the precision
deteriorates to $\sdmm=1.24\%$. The results are more or less symmetrical
around the maximal mixing. 
%Right panel shows the same information in
%relative precision $\sdmm$.

\begin{figure}[H]
\centering
\includegraphics[width=0.49\textwidth]
{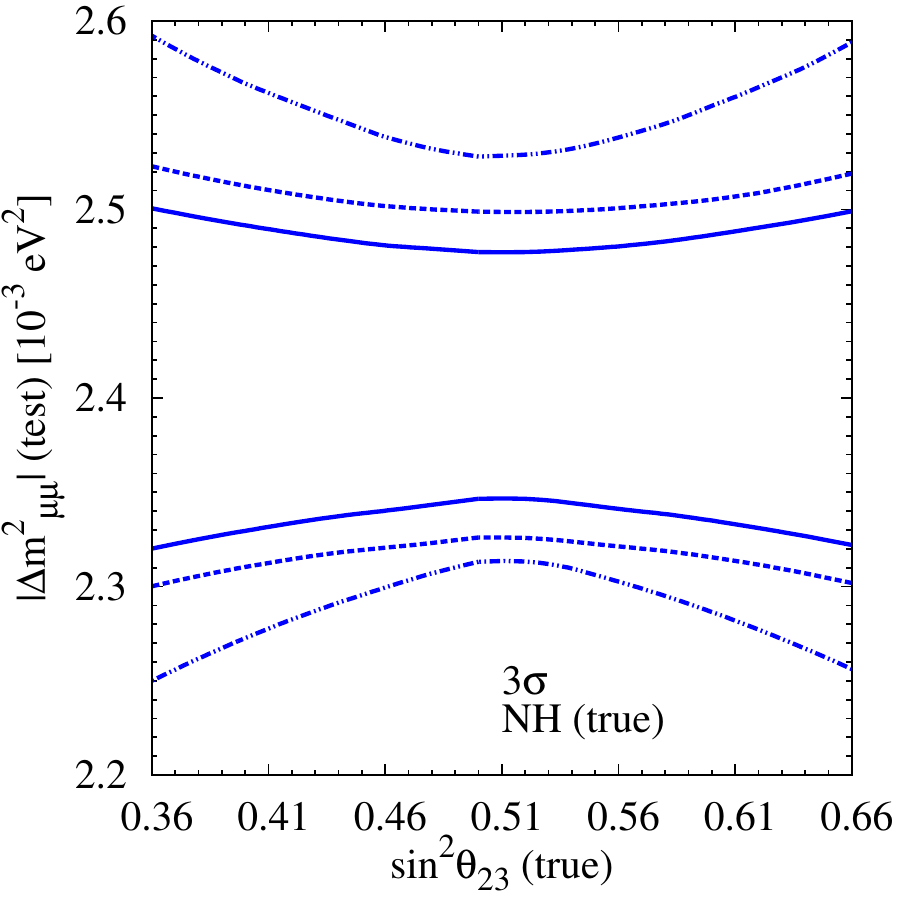}
\includegraphics[width=0.49\textwidth]
{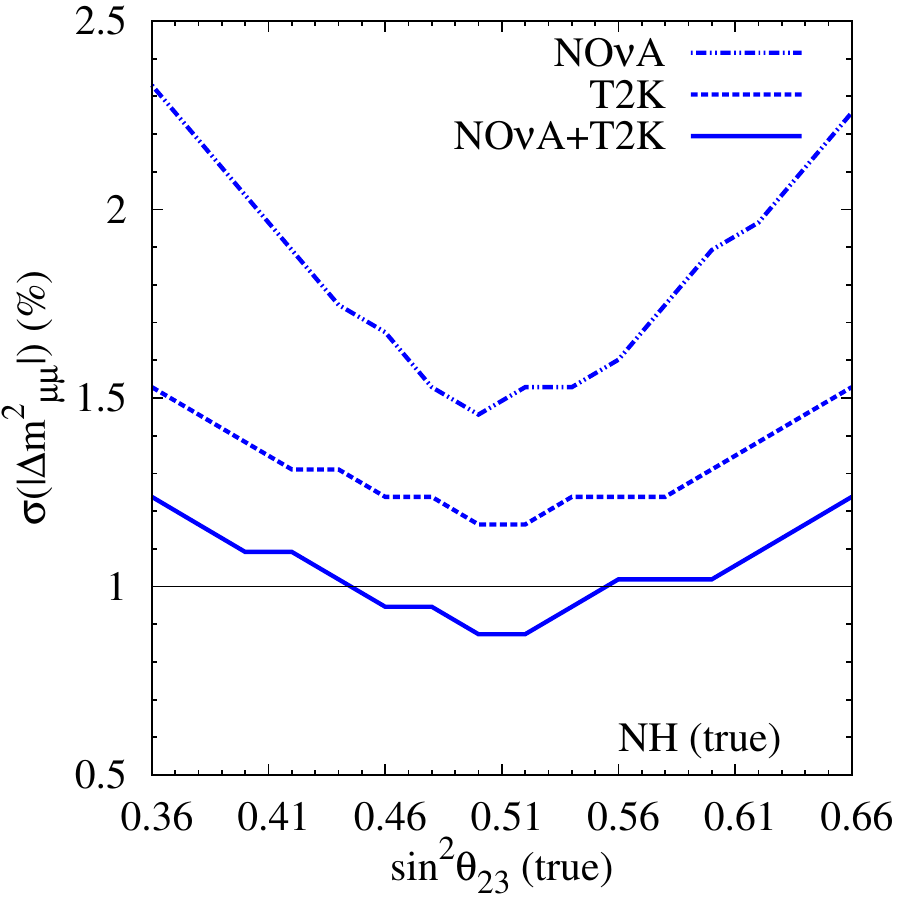}
\caption{\footnotesize{Left panel shows the 3\sig allowed regions in the
    $|\dmm|\text{(test)}$ - $\sin^2\tz\text{(true)}$ plane for the
    experiments T2K, NO$\nu$A, 
    and combined. Right panel depicts the corresponding relative 1$\sigma$
    precision on $|\dmm|$.
% as a function of $\sin^2\tz\text{(true)}$. 
    Here true hierarchy is NH, 
    $|\dmm|\text{(true)}=2.41\times10^{-3}~\text{eV}^2$,
    $\sin^22\ty\text{(true)}=0.089$, and $\dcp\text{(true)}=0$.}}
\label{dmm_theta23}
\end{figure}

\subsection{Effect of systematic uncertainties }
\label{systematics}

Here we study in detail the impact of the systematic
uncertainties on the measurement of $\dmm$. For this purpose,
we consider three different sets of assumptions on
systematics. The default choice of systematics has been already mentioned
in in Sec.~\ref{simulation}.
%These are 2.5\% signal normalization error and 10\%
%background normalization error. The normalization errors
%have been assumed to be the same for both the experiments. The energy scale
%uncertainty couples to $|\dmm|$ directly thus we expect it has affects to its
%precision. 
%We have taken a trivial linear assumption for the energy scale
%uncertainty, {\it i.e.}the energy scale uncertainty, named as ``tilt
%error'', generates a linear distortion to the survial spectrum\footnote{In
%  GLoBES, tilt error distorts the energy spectrum in the following way:
%  $E^{rec}_\nu = (1+\epsilon_\text{tilt})E_\nu^{true}$, where $E_\nu^{rec}$
%  and $E_\nu^{true}$ are the reconstructed and the true
%  neutrino/anti-neutrino energies and $\epsilon_\text{tilt}$ is the tilt
%  error.}. The tilt error is 0.01\% for NO$\nu$A for both signal and
%background, while it is 0.1\% for T2K for both signal
%and background. 
In the second set of systematic errors, we
increase the normalization error to 10\% and 20\% for both
signal and background, for both experiments, while keeping
the tilt errors same as before. In the third set, we further increase
the tilt error as well, to 10\% for both signal and backgrounds,
for both the experiments. The possible effect of these three
set of systematics on $\dmm$ precision is shown in table \ref{dmm_table}.
It can be seen that systematics play a minor role in the measurement
of $|\dmm|$.

%%%%%%%%%%%%%%%%%%%%%%%%%%%%%%%%%%%%%%%%%%%%%%%%%%%%%%%%
\begin{table}[H]
\begin{center}
\scalebox{0.9}{
\begin{tabular}{|c||c||c|c|c|}
\hline
NO$\nu$A
& Signal norm. err : Signal tilt err
& 2.5\% : 0.01\%
& 10\% : 0.01\%
& 10\% : 10\%
\\
& Bkg. norm. err : Bkg. tilt err
& 10\% : 0.01\%
& 20\% : 0.01\%
& 20\% : 10\%
\\
\hline
T2K
& Signal norm. err : Signal tilt err
& 2.5\% : 0.1\%
& 10\% : 0.1\%
& 10\% : 10\%
\\
& Bkg. norm. err : Bkg. tilt err
& 10\% : 0.1\%
& 20\% : 0.1\%
& 20\% : 10\%
\\

\hline\hline
\multicolumn{2}{|c||}{Relative precision $\sdmm$}
& 0.87\%
& 0.94\%
& 0.95\%
\\
\hline
\end{tabular}
}
\caption
{\footnotesize{Effect of systematic uncertainties on the relative 
1\sig precision of
    $|\dmm|$.}}
\label{dmm_table}
\end{center}
\end{table}
%%%%%%%%%%%%%%%%%%%%%%%%%%%%%%%%%%%%%%%%%%%%%%%%%%

%\subsection{Advancement in knowledge of precision with respect to the
%exposure}
\subsection{Evolution of $|\dmm|$ precision with statistics}

%Finally, in this section, we show how 
%the precision of $|\dmm|$ will improve with the accumulated statistics of
%the two experiments. 
%This is a safer quantity
%compared to the run-plan because of the uncertainties related to
%operation of the experiment itself. Thus, we do not correlate the 
%data available from T2K and NO$\nu$A individually, in a given
%period, but show the sensitivity achievable with increasing data.

\begin{figure}[H]
\centering
\includegraphics[width=0.49\textwidth]
{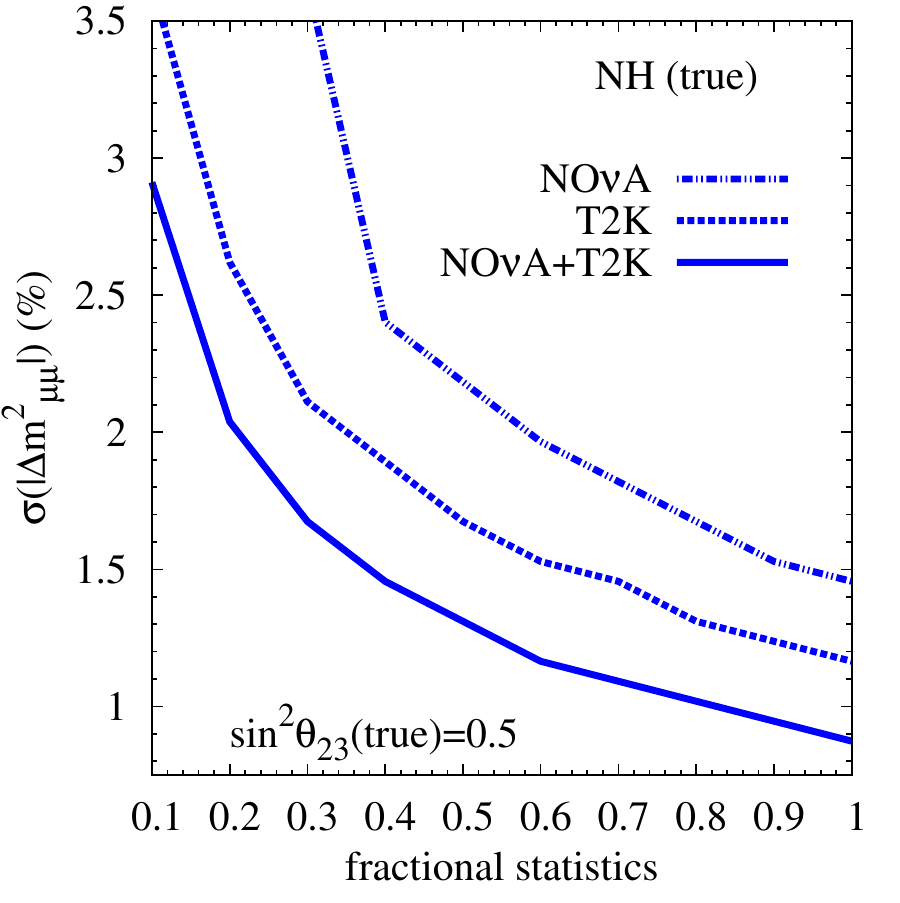}
\includegraphics[width=0.49\textwidth]
{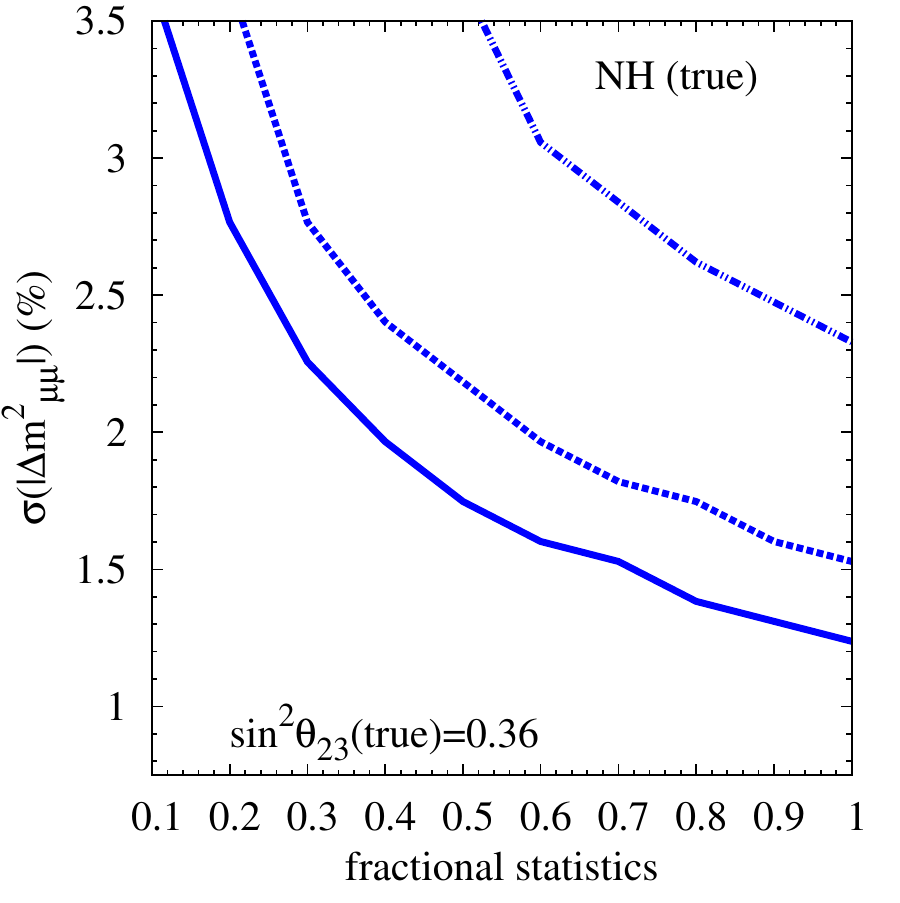}
\caption{\footnotesize{$\sdmm$ vs. fractional statistics for T2K,
    NO$\nu$A, and combined. Left (Right) panel corresponds to
  $\sin^2\tz\text{(true)}=0.5(0.36)$.
  Here true hierarchy is NH, $|\dmm|\text{(true)}=2.41\times10^{-3}$ eV$^2$,
  $\sin^22\ty\text{(true)}=0.089$, and $\dcp\text{(true)}=0$.}}
\label{dmm_expo}
\end{figure}

In Fig.~\ref{dmm_expo}, we study the improvement in
the precision of $|\dmm|$ as the statistics increases for T2K, NO$\nu$A, and
adding their data. 
% the evolution $\sdmm$ as a function
%of the fractional statistics in the unit of the total statistics based on
%their current run-plans. 
%In the x-axis, we show the fraction of the total
The x-axis shows the fraction of the total statistics for these experiments. 
%In producing Fig.~\ref{dmm_expo}, 
For NO$\nu$A, we assume equal run time in neutrino and anti-neutrino modes
at any given fractional statistics.
%We present the results for the most optimistic
%(0.5) and the most pessimistic (0.36) values of $\sin^2\tz\text{(true)}$.
Left and right panels present the results for the most optimistic
(0.5) and the most pessimistic (0.36) values of $\sin^2\tz\text{(true)}$
respectively.
% in regard with with $|\dmm|$ measurement. 
%An upper limit of $\sdmm\sim~3.5\%$
%is already available from the MINOS experiment. 
%We find that, for the most pessimistic assumption, on $\sin^2\tz$, 
%a $\sdmm=1.24\%$ is achievable with the combined data from these
%two experiments, corresponding to their full nominal-design statistics. 
For T2K, the precision improves from 3.5\% to 1.16\% as their statistics
increases from 10\% to 100\% for the maximal mixing case.
But, when we combine the data from T2K and NO$\nu$A with equal fractional
statistics, the precision improves from 2.9\% to 0.87\%.
%While, if $\sin^2\tz=0.5$,
%a $\sdmm=0.87\%$ can be achieved with the full exposure and 
A precision of $\sdmm=1\%$ can be achieved if 80\% of the total data from
the two experiments is available for maximal mixing.
%The combined data from T2K and 
%NO$\nu$A can bring down the present uncertainty of around 3.5\%
%in the determination of $|\dmm|$ to 1.5\% with only 40\% of their
%nominal design statistics. 
%As stand-alone experiments, the 
%best T2K and NO$\nu$A can achieve are $\sdmm=1.16\%$ and $\sdmm=1.45\%$
%respectively. This, of course, will happen if $\sin^2\tz=0.5$. If 
%$\sin^2\tz=0.36$, then their precisions will deteriorate to 1.52\%
%and 2.32\% respectively. 
For the pessimistic case (right panel), the precision on $|\dmm|$ improves
from 3.5\% to 1.24\% as the combined statistics increases from 10\% to
100\%.

{\it We have checked that a precision of 0.75\% is achievable with the
  combined data from T2K and NO$\nu$A 
  if their energy resolution can be improved by a factor of 2 assuming maximal
  2-3 mixing. We also would like to 
  point out that with the present energy resolution, the precision can be
  improved by simply increasing their statistics. 
  A precision of 0.61\% can be obtained if the statistics of these two
  experiments are doubled. It clearly suggests 
  that this measurement is still statistically dominated for the present
  run-plans of T2K and NO$\nu$A.
}

%{\it If the energy resolution of the detectors of T2K and
%NO$\nu$A improves by a factor of 2 
%100\%, 
%then, a precision of 
%0.75\% is achievable with the combined data for $\sin^2\tz\text{(true)}=0.5$.
%While, with the present resolutions, precision can be improved with
%increased statistics. We find that if the statistics of the two 
%experiments are doubled, a precision of 0.61\% is achievable
%with the combined data for $\sin^2\tz\text{(true)}=0.5$.}
%%%%%%%%%%%%%%%%%%%%%%%%%%%%%%%%%%%%%%%%%%%%%%%%%%%%%%%%%%%%%%%%%%%%%%%%%%%%%%%%%%%%%%%%%%%
%%%%%%%%%%%%%%%%%%%%%%%%%%%%%%%%%%%%%%%%%%%%%%%%%%%%%%%%%%%%%%%%%%%%%%%%%%%%%%%%%%%%%%%%%%%
%%%%%%%%%%%%%%%%%%%%%%%%%%%%%%%%%%%%%%%%%%%%%%%%%%%%%%%%%%%%%%%%%%%%%%%%%%%%%%%%%%%%%%%%%%%
\section{Summary and Conclusions\label{sec:summary}}

High-precision measurement of $|\dmm|$ is crucial in validating the 3-flavor
neutrino oscillation model. 
It also serves as a key input to the neutrino mass models and to the
neutrinoless double beta decay searches.
In addition, a sub-percent measurement of $|\dmm|$ is mandatory for the MBRO
experiments to obtain a reasonably
good sensitivity to neutrino MH. 
In the foreseeable future, presently running T2K and upcoming NO$\nu$A
experiments can provide a more accurate
measurement of $|\dmm|$ beyond the current MINOS precision.
In this paper, we have studied in detail the expected precision in $|\dmm|$
that can be achieved after the complete runs of
T2K and NO$\nu$A experiments.

\begin{table}[H]
\begin{center}
\scalebox{0.9}{
\begin{tabular}{|c||p{1.2in}|p{1.2in}|c|}
\hline
True $\sin^2\tz$
& \centering T2K (5$\nu$)
& \centering NO$\nu$A (3$\nu$ + 3$\anu$)
& T2K + NO$\nu$A
\\
\hline
0.36
& \centering 1.53\%
& \centering 2.33\%
& 1.24\% ($2.41^{+0.09}_{-0.09}$)
\\
\hline
0.50
& \centering 1.16\%
& \centering 1.45\%
& {\bf 0.87\%} ($2.41^{+0.07}_{-0.06}$)
\\
\hline
0.66
& \centering 1.53\%
& \centering 2.26\%
& 1.24\% ($2.41^{+0.09}_{-0.09}$)
\\
\hline
\end{tabular}
}
\caption{
\footnotesize{Relative 1$\sigma$ precision on $|\dmm|$ considering
    different true values of $\sin^2\tz$. Results are shown for T2K,
    NO$\nu$A, and their combined data. In the last column, inside the
    parentheses, we also give the 3$\sigma$ allowed ranges of test $|\dmm|$
    ($\times10^{-3}~\text{eV}^2$) around its best-fit.}}
%theta23 values for T2K, NOvA, and T2K+NO$\nu$A }
%Expected $|\dmm|$ precisions for different $\sin^2\tz \text{(true)}$ values
%  using  different datasets}
\label{tab_results}
\end{center}
\end{table}
%%%%%%%%%%%%%%%%%%%%%%%%%%%%%%%%%%%%%%%%%%%%%%%%%%

%In Table~\ref{tab_results}, we show the 
%3\sig allowed range of test $|\dmm|$ and the precision $\sdmm$,
%respectively, for different true values of $\sin^2\tz$.
It can be seen from Table~\ref{tab_results} that T2K (NO$\nu$A) 
can measure $|\dmm|$ with a relative 1$\sigma$ precision of
1.45\% (1.16\%) assuming maximal 2-3 mixing. Combining the data 
from these two experiments, a sub-percent precision is achievable.
It clearly demonstrates the possible synergy between these two 
experiments with different energy spectra and baselines.
%We find that with the combined data, a sub-percent precision is achievable, 
%provided the 2-3 mixing is close to maximal. 
%These experiments,
%individually, will not provide a sub-percent precision. 
%Our findings are summarized in Table~\ref{tab_results}.
We have also studied the dependency of this measurement on the true value 
of $\sin^2\tz$. The precision in $|\dmm|$ can vary in the range of 0.87\% to
1.24\% depending on the true value of $\sin^2\tz$ in its currently-allowed
3\sig region. As expected, for maximal 2-3 mixing, we have the best
measurement of  0.87\% (see Table~\ref{tab_results}). 
Any analysis assuming the full runs of these two long-baseline experiments
can now assume a 1$\sigma$ prior of $\sim1$\% on $|\dmm|$.
In the last column,
inside the parentheses, we also present the 3$\sigma$ allowed ranges of test
$|\dmm|$ ($\times10^{-3}~\text{eV}^2$) around its best-fit.
This is a very robust measurement in the sense that it is quite immune to
the present  uncertainties in $\sin^22\ty$, $\dcp$, choice of hierarchy, and
the systematic errors. 
This high-precision measurement of $|\dmm|$ by the
current generation experiments T2K and NO$\nu$A will certainly provide a
boost for the physics reach of MBRO 
experiments in addressing the neutrino mass hierarchy.

%the precision in $|\dmm|$ will vary from 0.87\% to 1.23\%. 

%We further demonstrate that this is a robust measurement as it remains almost
%unaffected by the present uncertainties in $\ty$, $\dcp$, the choice of mass
%hierarchy, and the systematic errors.

%demonstrated that this measurement is highly dependent
%on the true value of $\sin^22\tz$ and the best precision will
%be achieved if 2-3 mixing is maximal. 
%This will be a very robust measurement as it is immune to the present uncertainties in
%$\sin^22\ty$, hierarchy, $\dcp$ and the systematic errors. 
%We hope such a
%stable precise measurement on $|\dmm|$ by combining T2K and NO$\nu$A data
%could provide a reference point to the physics case of medium-baseline
%reactor neutrino experiments aimed at resolving neutrino MH.

\subsubsection*{Acknowledgments}
We would like to thank Jun Cao, Yu-Feng Li, Eligio Lisi, Panagiotis Stamoulis, 
and Yifang Wang for useful discussions. SKA acknowledges the support from DST/INSPIRE Research Grant
[IFA-PH-12], Department of Science and Technology, India. SP acknowledges
support from the Neutrino Project under the XII plan of Harish-Chandra
Research Institute.

%%%%%%%%%%%%%%%%%%%%%%%%%%%%%%%%%%%%%%%%%%%%%%%%%%%%%%%%%%%%%%%%%%%%%%%%%%%%%%%%%%%%%%%%%%
%%%%%%%%%%%%%%%%%%%%%%%%%%%%%%%%%%%%%%%%%%%%%%%%%%%%%%%%%%%%%%%%%%%%%%%%%%%%%%%%%%%%%%%%%%%
%%%%%%%%%%%%%%%%%%%%%%%%%%%%%%%%%%%%%%%%%%%%%%%%%%%%%%%%%%%%%%%%%%%%%%%%%%%%%%%%%%%%%%%%%%%

\begin{appendix}

%\section{Effect of the variation in true $\dcp$}
%\label{appB}

%Here, we demonstrate the fact that the sensitivities of the experiments 
%T2K and NO$\nu$A for the determination of $\dmm$ is nearly-independent
%of the assumed true value of $\dcp$. We have shown the results in Fig.
%\ref{dcp_variation} where we show the $\dxx$ as a function of 
%test $\dmm$ for various different true values of $\dcp$. The left (right)
%panel is assuming the true hierarchy to be NH (IH).

%\begin{figure}[H]
%\centering
%\includegraphics[width=0.49\textwidth]
%{./Figures/delta31precision_nova33t2k50_NH_3sig_ssqt23_0pt50.pdf}
%\includegraphics[width=0.49\textwidth]
%{./Figures/delta31precision_nova33t2k50_IH_3sig_ssqt23_0pt50.pdf}
%\caption{$\dxx$ vs. test $\dmm$ for the experiment NO$\nu$A. The 
%left (right) panel corresponds to NH (IH) being the true hierarchy.
%The effect of variation in the value of $\dcp$ has been shown.
%Here, true $\dmm=2.41\times10^{-3}\text{eV}^2$, 
%true $\sin^2\tz=0.5$ and true $\sin^22\ty=0.089$.}
%\label{dcp_variation}
%\end{figure}

\section{Impact of 1-3 mixing angle on $|\Delta m^2_{\mu \mu}|$ precision}
\label{appA}

The $\nu_{\mu}\rightarrow\nu_{\mu}$ survival probability is independent of $\ty$ to
the first order. 
%Though the present 3\sig confidence level
%limit of $\sin^22\ty$, [0.067, 0.111], is quite sizable, 
%Thus, we do not expect this
%crucial parameter for the appearance channel has 
%significant effect 
%much impact on $\sdmm$ which is dominated by the disappearance channel. 
%This is shown clearly in 
Therefore, the precision measurement of $|\dmm|$ should not be
affected much by this parameter.
This is indeed the case as shown in Fig.~\ref{dmm_theta13}
where we have presented the precisions of $|\dmm|$ for the best-fit as
well as the 3\sig upper and lower limits of true $\sin^22\ty$. The
$|\dmm|$ precision achieved by the experiment NO$\nu$A is not
affected by the uncertainty in $\sin^22\ty$. The same is true
for T2K.
%will withstand the current uncertainty in $\sin^22\ty$.

\begin{figure}[H]
\centering
\includegraphics[width=0.49\textwidth]
{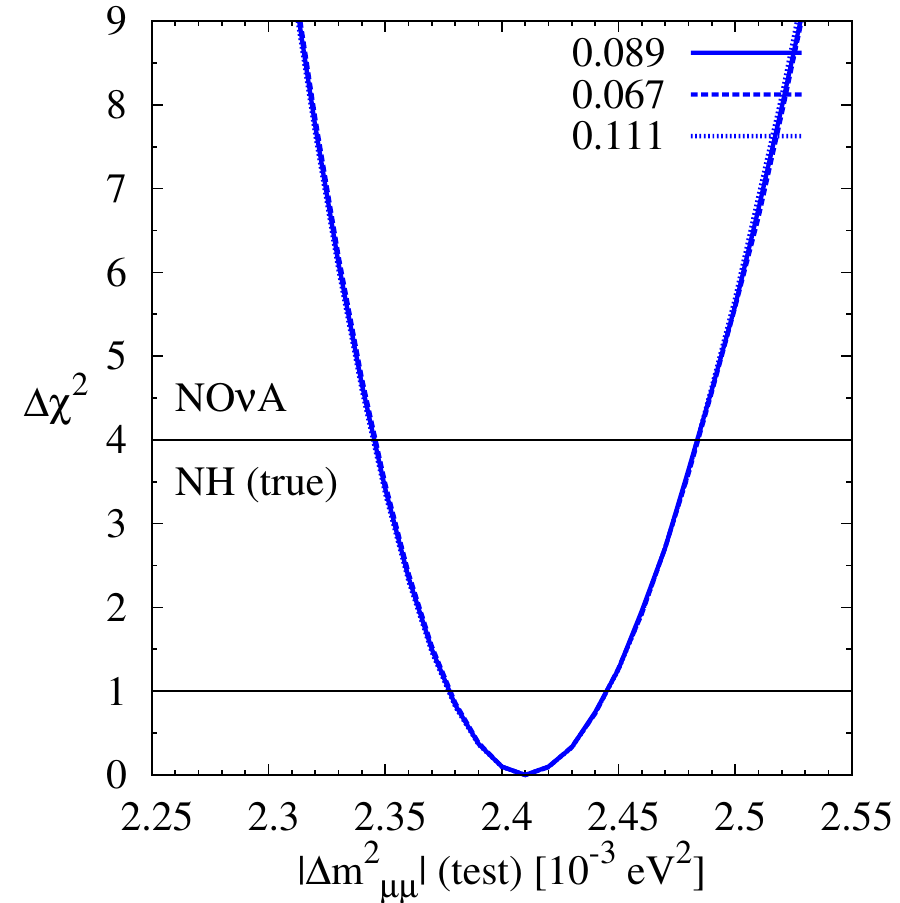}
\includegraphics[width=0.49\textwidth]
{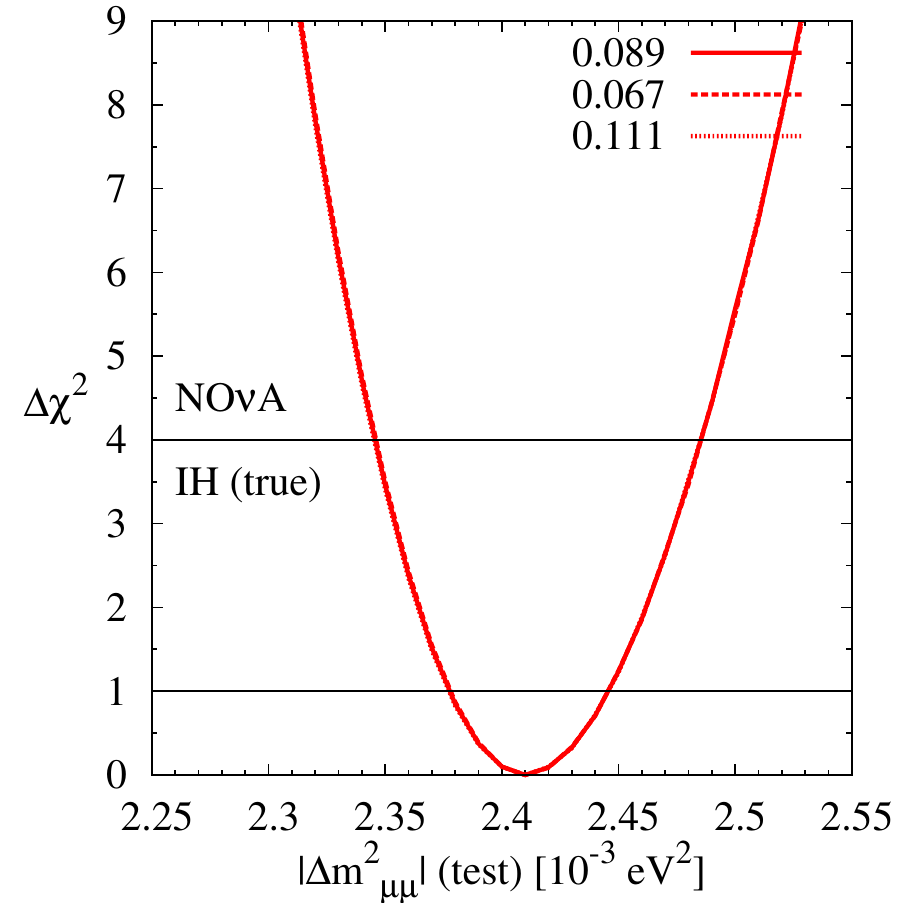}
\caption{\footnotesize{$\dxx$ vs. $|\dmm|\text{(test)}$ for NO$\nu$A. Left
(right) panel corresponds to true NH (IH).
The effect of $\sin^22\ty$ on the precision of $|\dmm|$ has been shown for
three different true $\sin^22\ty$ values: the current best-fit and the 3\sig
upper and lower limits.
Here $|\dmm|\text{(true)}=2.41\times10^{-3}~\text{eV}^2$, 
$\sin^2\tz\text{(true)}=0.5$, and $\dcp\text{(true)}=0$.}}
\label{dmm_theta13}
\end{figure}

\end{appendix}

\bibliographystyle{apsrev}
\bibliography{references}

\end{document}